\documentstyle[prl,aps,epsf]{revtex}
\begin{document} \draft \preprint{ISSP-96-1-11}
\twocolumn[\hsize\textwidth\columnwidth\hsize\csname @twocolumnfalse\endcsname
\title{Universal low-temperature properties of quantum and classical\\
ferromagnetic chains}
\author{Minoru Takahashi$^a$, Hiroaki Nakamura$^a$, and
Subir Sachdev$^{b}$}
\address{$^a$ Institute for Solid State Physics,
University of Tokyo, Roppongi, Minato-ku, Tokyo 106, Japan}
\address{$^b$ Department of Physics, P.O. Box 208120, Yale University,
New Haven, CT 06520-8120, USA}
\maketitle
\begin{abstract}
We identify the critical theory controlling the universal, low temperature,
macroscopic properties of both quantum and classical ferromagnetic chains.
The theory is the quantum mechanics of a single rotor. The mapping
leads to an efficient method for computing scaling functions to high
accuracy.
\end{abstract}
\vskip2pc] \narrowtext

A number of recent papers~\cite{Takyam1,Takyam2,naka1,naka2,rs}
have studied the
finite temperature properties of  ferromagnetic quantum spin chains. At low
temperatures, macroscopic observables  can be fully 
described 
by two dimensionful parameters which characterize the ground state. 
A convenient choice for these is the
ground state magnetization density $M_0$, and the ground state spin stiffness
$\rho_s$. Then the macroscopic properties of a quantum spin chain of length $L$,
in the presence of an external magnetic field $H$, and at a temperature $T$ are
fully universal functions of the dimensionless ratios that can be formed out of
these parameters. A convenient choice for these ratios is (recall that in $d=1$
$\rho_s$, has dimensions of energy $\cdot$ length)
\begin{equation} r \equiv \rho_s M_0 /T,~~~h \equiv H/T,~~~
q \equiv \rho_s/(L T).
\end{equation}
Here we have used units in which $\hbar = k_B = 1$ and absorbed a factor of $g
\mu_B$ into $H$ ($\mu_B$ is the Bohr magneton).
Thus, for instance, the temperature dependent magnetization density
obeys~\cite{rs}
$M = M_0 \Phi_M ( r,h,q)$ where $\Phi_M$ is a universal function. This, and
other, universal functions will depend upon the boundary conditions on the
chain: we will focus, for simplicity, on periodic boundary
conditions {\em i.e.} on spin rings. There is no method for computing
$\Phi_M$ in complete generality---in this paper we shall show how to compute
$\Phi_M$ efficiently in a limiting case when the quantum ring can be described
by an effective classical model.

A convenient point for beginning discussion is the spin-wave expansion. This
expansion is valid provided $r \gg 1$ (provided $h$ is not too small), and it is
quite easy to use standard methods to determine the leading term.
For a ring we find
\begin{equation}
\Phi_M (r,h,q) = 1 - \frac{q}{r} \sum_{n} \frac{1}{\exp(4 \pi^2 n^2 q^2 / r +
h) - 1} \ldots ,
\end{equation}
where the sum is over all integers $n$. An interesting property of this
expression emerges in the limit $r \rightarrow \infty$, $h \rightarrow 0$, but
with
\begin{equation}
g \equiv r h = \rho_s M_0 H / T^2,
\end{equation}
fixed; then we find
$\Phi_M = 1 - (1/2\sqrt{g}) \coth (\sqrt{g}/2q)$. The implication of recent
works~\cite{naka1,naka2,rs} is that this limit is non-trivial at each order
in the spin-wave expansion, and that the resulting series has in 
fact properties of the {\em classical\/} ferromagnetic ring. 
Thus we may define the classical scaling function $\phi_M$ by
\begin{equation}
\phi_M (g, q) = \lim_{r \rightarrow \infty} \Phi_M ( r, g/r, q).
\end{equation}
Classical behavior emerges in this limit because the ferromagnetic correlation
length becomes larger than the de Broglie wavelength of the spin waves.

One possible approach to the computation of the scaling function $\phi_M$ is to
compute the magnetization of a nearest-neighbor, classical ferromagnetic chain,
whose statistical mechanical properties were computed 
some time ago~\cite{tuto52,fish64,joyce67}. 
The scaling limit of classical solution 
was studied in recent work~\cite{naka1,naka2}, and led {\em e.g.} to the result 
$\phi_M(g, 0) ={2\over 3}g- {44\over 135}g^3 + {\cal O} (g^5)$
---this result means that the usual linear susceptibility 
$\partial M/\partial H$  diverges as $T^{-2}$ and that the third
order non-linear susceptibility
$\partial^3 M/\partial H^3$ diverges as $T^{-6}$.
However, the computations required to
achieve this limited result were quite 
complicated. 
In this paper we shall develop an efficient method to computing 
the complete function $\phi_M (g,q)$ to essentially arbitrary accuracy. 
This will be done by a precise identification of the critical theory 
controlling this crossover function.

We begin by considering the partition function of a
classical Heisenberg model in a uniform magnetic field:
\begin{displaymath}
Z=\int\exp\Bigl(\sum_{i<j}{J(i-j)\over T}{\bf n}_i \cdot {\bf n}_j
+{H\over T}\sum_{i=1}^N n^z_i\Bigr)\prod_i
d{\bf n}_i,
\end{displaymath}
where the exchange constants $J(i) \geq 0$, $J(i) = J(N-i)$, the ${\bf n}_i$
are unit 3-component vectors which are integrated over.
The universal critical theory
emerges when we take the continuum limit of this action. With a lattice spacing
$a$, the continuum limit will be characterized by the values $M_0 = 1/a$,
$L = Na$, and
$\rho_s = a \sum_{i} i^2 J(i)$. Our results apply to all systems in
which the summation in the definition of $\rho_s$ is convergent---for
$J(i ) \sim i^{-q}$ this is the case for $q > 3$.
The continuum field theory which emerges by this method gives the
partition function $Z_c$:
\begin{displaymath}
\int {\cal D}[{\bf n}]
\exp\left(- \int^L_0 
{\rho_s \over 2T}\left({d {\bf n}(x)\over d x}\right)^2 -
{H M_0 \over T}
n^z(x) dx \right),
\end{displaymath}
where the integral is now a functional integral over  unit vector fields ${\bf
n} (x)$ satisfying ${\bf n} (0) = {\bf n} (L)$.
A key property of $Z_c$ is that it is a finite field theory, free of ultraviolet
divergences. This becomes clear when we re-interpret $Z_c$ as the
imaginary ``time'' ($x$) Feynman path integral for the quantum mechanics of a
single particle with co-ordinate ${\bf n} (x)$ restricted to lie on a unit
sphere: no discretization of time is required to define this problem.
As a result, all observables are universal functions of the couplings in $Z_c$,
and it is useful to now transform to dimensionless variables.
We rescale spatial co-ordinates
$y=Tx/\rho_s$, and obtain $Z_c$:
\begin{displaymath}
\int{\cal D}[{\bf n}]\exp\left(- \int^{1/q}_0 
{1\over 2}\left({d {\bf n}(y)\over d y}\right)^2 -g n^z(y)dy \right).
\end{displaymath}
\par
Subsequent computations are best carried out using the Hamiltonian, ${\cal H}$,
of the quantum particle described by $Z_c$:
\begin{equation}
{\cal H}={{\bf L}^2\over 2}-gn^z.\label{eq:rotor}
\end{equation}
This describes a single {\em quantum rotor} with unit moment of inertia, angular
momentum operator ${\bf L}$ (which obeys the usual commutation relations
$[ L^{\alpha}, L^{\beta} ] = i \epsilon_{\alpha\beta\gamma} L^{\gamma}$),
in the presence of a ``gravitational'' field $g$. There is no need to consider
the radial motion as the length of ${\bf n}$ is constrained to unity. The
logarithm of $Z_c$ equals the free energy of the quantum system ${\cal H}$ at a
``temperature'' $q$; in the original spin ring, $q$ is the ratio of
correlation length at $H=0$ to length of the system. For other boundary
conditions, $Z_c$ will be given by appropriate propagators of ${\cal H}$.\par

We now consider eigenvalue equation ${\cal H} \psi = E \psi$.
As ${\cal H}$ commutes with $L^z$, eigenstates are divided to subspaces
of azimuthal quantum number $m=...-3,$ $-2,-1,0,1,2,3,...$. In spherical
co-ordinates $(\theta, \varphi)$, $\psi$ is given by $e^{im \varphi}u(\theta)$.
At $g=0$ $u(\theta)$ is given by the associated Legendre Polynomials 
$P^{\vert m\vert}_l(\cos\theta)$ and eigenstates
of
${\cal H}$ are spherical harmonic states $|l,m\rangle$ with $l\ge |m|$.
The matrix elements of ${\cal H}$ in this basis are
\begin{eqnarray}
&&<l',m\vert {\cal H}\vert l,m>={l(l+1)\over 2}\delta_{ll'} \nonumber \\
&& -g\Bigl(\delta_{l,l'+1}\sqrt{l^2-m^2\over 4l^2-1}+
\delta_{l',l+1}\sqrt{l'^2-m^2\over 4l'^2-1}\Bigr).\label{eq:rotor2}
\end{eqnarray}
Notice that ${\cal H}$ is tri-diagonal in each $m$ subspace:
this makes numerical diagonalization of ${\cal H}$ quite
straightforward.
The eigenvalues of Hamiltonian are given by $E_{m,n}(g)$, and
$n=0,1,2,...$ is the number of nodes of function $u(\theta)$.
We also generated a power series expansion in $g$ for the ground state energy
$E_{0,0}(g)$ using a symbolic manipulation program (Mathematica); this leads
to the magnetization scaling function for the infinite ferromagnetic ring
$\phi_M (g, 0) = -dE_{0,0}(g)/dg$, for which we find
\begin{eqnarray}
&&\phi_M (g,0)={2\over 3}g-{44\over 135}g^3+{752\over 2835}g^5
-{465704\over 1913625}g^7\nonumber \\
&&+{356656\over 1515591}g^9
-{707126486624\over 3016973334375}g^{11}+{1126858624\over 4736221875}g^{13}
\nonumber\\
&&
-{ 5083735857217648\over 20771861407171875 }g^{15}+....
\label{eq:expan1}
\end{eqnarray}
All coefficients are rational numbers.

In the complementary large $g$ limit, the particle spends most of its time near
the `north pole', and in its vicinity it experiences a harmonic oscillator
potential well. It therefore pays to work now in the basis states of this
harmonic oscillator and thereby generate a perturbation
expansion valid for large $g$.
Parametrizing $E = -g + \sqrt{g}\varepsilon$
and
$u(\theta)=(\theta/\sin(\theta))^{1/2} f ( g^{1/4} \theta)$
; this gives us an eigenvalue equation for $f$: $( h_0 + h_1 ) f(z)
= \varepsilon f(z)$ with
\begin{displaymath}
h_0 = \frac{1}{2} \left( - \frac{1}{z} \frac{d}{dz}  z \frac{d}{dz}
+ \frac{m^2}{z^2} + z^2 \right),\quad h_1=
\end{displaymath}
\begin{displaymath}
2 \sqrt{g} \left( \sin^2 \frac{\theta}{2} - \frac{\theta^2}{4} \right)
+ \frac{1}{8 \sqrt{g}} \left( \frac{1-4m^2}{\theta^2} -
\frac{1-4m^2}{\sin^2 \theta}
- 1 \right),
\end{displaymath}
where
$z \equiv  g^{1/4}\theta$.
Notice that $h_0$ describes a two-dimensional harmonic oscillator in radial
co-ordinates. Its eigenstates $|n\rangle,~~n \geq 0$ 
have energy $\varepsilon_0= 2n + |m| + 1$ and 
are represented by generalized Laguerre polynomials 
$z^{|m|}L^{|m|}_{n+|m|}(z^2)e^{-z^2/2}$. 
Further, notice that $h_1$ can be
expanded as a series in positive integer powers of $z^2$, with all terms being
small for large $g$. The matrix elements of $h_1$ in the $|n\rangle$ basis can
be determined by repeated use of the identity
$z^2 |n \rangle = (2n + |m| + 1)|n \rangle$ $- \sqrt{n(n+|m|)} |n-1\rangle
- \sqrt{(n+1)(n+|m|+ 1)} | n+1\rangle$. It now remains to diagonalize $h$ in the
$|n\rangle$ basis, which can be done order by order in $g^{-1/2}$ by
Mathematica.
Such a procedure was used to generate an expansion for the ground state energy,
$E_{0,0}(g)$ and hence for $\phi_M (g, 0 )$:
\begin{eqnarray}
&&\phi_M (g, 0) =1-{g^{-1/2}\over 2}-{g^{-3/2}\over 128}-{3 g^{-2}\over 512}
-{159g^{-5/2}\over 32768} \nonumber \\
&& -{297g^{-3}\over 65536}-{19805g^{-7/2}  \over
4194304}
-{91089g^{-4} \over 16777216 }-{ 14668507g^{-9/2}\over 2147483648}
\nonumber \\
&&-{ 20057205 g^{-5}\over 2147483648}
-{3794803731 g^{-11/2} \over 274877906944}-....
\label{eq:expan2}
\end{eqnarray}
Unlike the small $g$ expansion, which has a finite radius of convergence,
the large $g$ expansion is only asymptotic. In particular, the large $g$ limit
loses topological information associated with tunneling paths which traverse the
south pole---such paths will lead to `instanton' contributions which are
exponentially small for large $g$. In Table 1 we give the higher
order coefficients of these expansions.

For the quantum ferromagnetic Heisenberg ring with spin $1/2$:
\begin{eqnarray}
&&{\cal H}=-\sum_{ k=1}^NJ{\bf S}_k{\bf S}_{k+1}
-H\sum_{k=1}^N S_k^z, \nonumber \\
&&[S_l^\alpha,S_k^\beta ]=i\delta_{lk}\epsilon_{\alpha\beta\gamma}S_l^\gamma,
\label{eq:quham}
\end{eqnarray}
we can calculate the magnetization at $N=\infty$ limit
for given temperature and magnetic field
using thermodynamic Bethe ansatz equations\cite{naka1}.
The magnetization still obeys the same limiting scaling function.
The stiffness constant $\rho_s$ is $Ja/4$ and $M_0$ is $1/2a$.
In Fig.2 we compare the magnetization as a function of $g=JH/8T^2$.
As $T$ goes down, the magnetization divided by $M_0$
approaches the scaling function $\phi_M(g,0)$, supporting the conclusion
that the quantum ferromagnetic Heisenberg ring has the same magnetic
scaling function with classical one. \par

In our scaling theory, the linear susceptibility $\partial M/\partial H$
of quantum ferromagnet diverges as $T^{-2}$ at low temperature.
Schlottmann\cite{schlot} proposed the divergence of the type $T^{-2}/\ln(J/T)$
using the numerical analysis of thermodynamic Bethe ansatz equations.
But later this was negated by the detailed investigations of thermodynamic
Bethe ansatz equations\cite{taka1}.

Next we consider the function $\phi_M(g,q)$ at $q>0$, which is important
for the analysis of short rings.
This is represented by the eigenvalues
$E_{m,n}(g)$ of Hamiltonian (\ref{eq:rotor}):
\begin{equation}
\phi_M(g,q)=-{\sum_m\sum_n E_{m,n}'(g)\exp(-E_{m,n}(g)/q)\over
\sum_m\sum_n \exp(-E_{m,n}(g)/q)}.
\end{equation}
For $q\ll 1$, $\phi_M$ is dominated by the ground state $m=n=0$. The energy
gap to the second lowest eigenvalue at $m=\pm 1, \quad n=0$ is more than 1.
Then, deviations from $\phi_M(g,0)$ are exponentially small:
\begin{equation}
\phi_M(g,q)=\phi_M(g,0)+{\cal O}(e^{-1/q}),\quad q \ll 1.
\end{equation}
In Table 2 we give the result of numerical calculation of $\phi_M(g,q)$.
We calculate $E_{m,n}(g)$ and $E'_{m,n}(g)$ numerically by diagonalizing
the tridiagonal matrices (\ref{eq:rotor2}). Terms at very big $n$ or
$m$ are not necessary because their contributions are exponentially small.
\par

Finally, we note that the behavior of the scaling functions is also simple in
the limit
$q\gg 1$. The problem is now equivalent to a single {\em classical\/} rotor:
\begin{equation}
\phi_M(g,q)=\coth{g\over q}-{q\over g}, \quad {g\over q}={M_0HL\over T}.
\end{equation}
This means that the system behaves as one big spin $M_0L$ if the
correlation length $\rho_s/T$ is much longer than the system size $L$.

This research is supported in part by
Grants-in-Aid for Scientific Research on Priority Areas,
``Infinite Analysis'' (Area No 228),
from the Ministry of Education, Science and Culture, Japan,
and by the U.S. National Science Foundation Grant No  DMR-92-24290.

\begin{table}
\caption{Expansion coefficients of $\phi_M(g,0)$ for small g expansion
$\sum a_ng^{2n-1}$ and large $g$ expansion $1-{1\over 2}g^{-1/2}$
$+\sum b_ng^{-(n+2)/2}$. }
\begin{tabular}{rll}
$n$ &  $a_n$                 & $b_n$                 \\
\tableline
1&     0.66666666666666667   & -0.0078125            \\
2&    -0.32592592592592593   & -0.005859375          \\
3&     0.265255731922398589  & -0.004852294921875    \\
4&    -0.243362205238748449  & -0.0045318603515625   \\
5&     0.235324701717019961  & -0.0047218799591064453\\
6&    -0.234382743316652222  & -0.0054293274879455566\\
7&     0.237923529289894173  & -0.0068305558525025845\\
8&    -0.24474146816049874   & -0.0093398639000952244\\
9&     0.25422990405144852   & -0.0138054155504505616\\
10&   -0.266079264257003861  & -0.02195840459989995  \\
11&    0.280145597828282407  & -0.037433421774196063 \\
12&   -0.296386895761683996  & -0.068149719593456837 \\
13&    0.314829857750047932  & -0.132063536806409254 \\
14&   -0.33555167675962067   & -0.271575442615252266 \\
15&    0.35866981723790861   &                       \\
16&   -0.38433633684212508   &                       \\
17&    0.41273495146936857   &                       \\
18&   -0.44407985807794498   &                       \\
19&    0.47861575559844195   &
\end{tabular}
\end{table}
\begin{table}
\caption{Values of scaling function $\phi_M(g,q)$} for $q=0.0,0.5,1.0,1.5,2.0$
\begin{tabular}{ccccccr}
$g\backslash q$&0&0.5&1.0&1.5&2.0\\
\tableline
 0.0 &   0.0      &   0.0      &   0.0      &    0.0     &   0.0     \\
 0.2 &   0.130808 &   0.093330 &   0.056126 &   0.039675 &   0.030627\\
 0.4 &   0.248178 &   0.182802 &   0.111511 &   0.079100 &   0.061141\\
 0.6 &   0.345162 &   0.265293 &   0.165461 &   0.118034 &   0.091434\\
 0.8 &   0.421694 &   0.338855 &   0.217367 &   0.156249 &   0.121399\\
 1.0 &   0.481193 &   0.402773 &   0.266734 &   0.193537 &   0.150934\\
 1.2 &   0.527688 &   0.457313 &   0.313199 &   0.229716 &   0.179947\\
 1.4 &   0.564580 &   0.503351 &   0.356529 &   0.264633 &   0.208353\\
 1.6 &   0.594419 &   0.542043 &   0.396618 &   0.298164 &   0.236074\\
 1.8 &   0.619027 &   0.574582 &   0.433462 &   0.330217 &   0.263046\\
 2.0 &   0.639690 &   0.602067 &   0.467146 &   0.360730 &   0.289213\\
 2.2 &   0.657323 &   0.625439 &   0.497818 &   0.389667 &   0.314529\\
 2.4 &   0.672584 &   0.645479 &   0.525667 &   0.417020 &   0.338961\\
 2.6 &   0.685956 &   0.662813 &   0.550909 &   0.442800 &   0.362483\\
 2.8 &   0.697795 &   0.677939 &   0.573769 &   0.467038 &   0.385080\\
 3.0 &   0.708374 &   0.691253 &   0.594470 &   0.489780 &   0.406745\\
 3.2 &   0.717902 &   0.703067 &   0.613231 &   0.511084 &   0.427481\\
 3.4 &   0.726544 &   0.713628 &   0.630253 &   0.531013 &   0.447294\\
 3.6 &   0.734429 &   0.723134 &   0.645724 &   0.549640 &   0.466199\\
 3.8 &   0.741664 &   0.731744 &   0.659815 &   0.567040 &   0.484216\\
 4.0 &   0.748332 &   0.739587 &   0.672678 &   0.583287 &   0.501368\\
 4.2 &   0.754506 &   0.746768 &   0.684450 &   0.598457 &   0.517682\\
 4.4 &   0.760244 &   0.753373 &   0.695250 &   0.612624 &   0.533188\\
 4.6 &   0.765594 &   0.759476 &   0.705186 &   0.625859 &   0.547918\\
 4.8 &   0.770600 &   0.765134 &   0.714351 &   0.638232 &   0.561904\\
 5.0 &   0.775297 &   0.770401 &   0.722827 &   0.649805 &   0.575181\\
 5.2 &   0.779715 &   0.775319 &   0.730686 &   0.660640 &   0.587781\\
 5.4 &   0.783881 &   0.779924 &   0.737991 &   0.670794 &   0.599739\\
 5.6 &   0.787819 &   0.784248 &   0.744798 &   0.680319 &   0.611089\\
 5.8 &   0.791548 &   0.788320 &   0.751156 &   0.689265 &   0.621862\\
 6.0 &   0.795087 &   0.792162 &   0.757108 &   0.697676 &   0.632091

\end{tabular}
\end{table}
\begin{figure}
\epsfxsize=8cm \epsffile{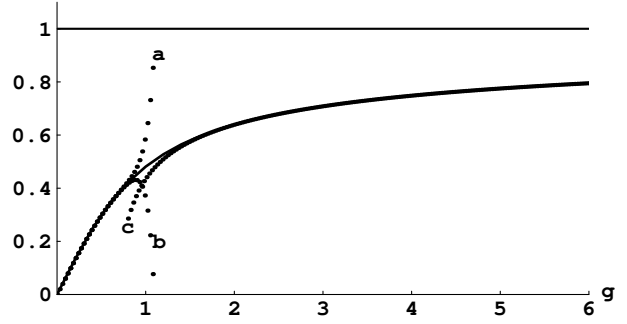}
\caption{
The scaling function $\phi_M(g,0)$. Line a is expansion
(\protect\ref{eq:expan1})
up to $g^{13}$ and Line b is expansion up to $g^{15}$. Line c
is the result of asymptotic expansion
(\protect\ref{eq:expan2}) up to $g^{-11/2}$ from $g=\infty$.
}
\end{figure}
\begin{figure}
\epsfxsize=8cm \epsffile{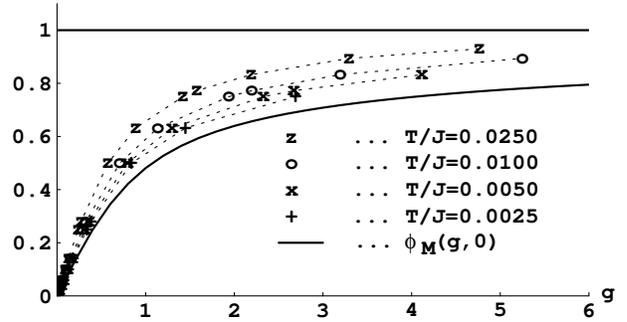}
\caption{Magetization versus $g=JH/8T^2$ for spin-half and infinite-length
ferromagnetic Heisenberg chain with nearest neighbor
exchange (\protect\ref{eq:quham}).
As temperature goes down, the line approaches to the theoretical
line $\phi_M(g,0)$.}
\end{figure}
\begin{figure}
\epsfxsize=8cm \epsffile{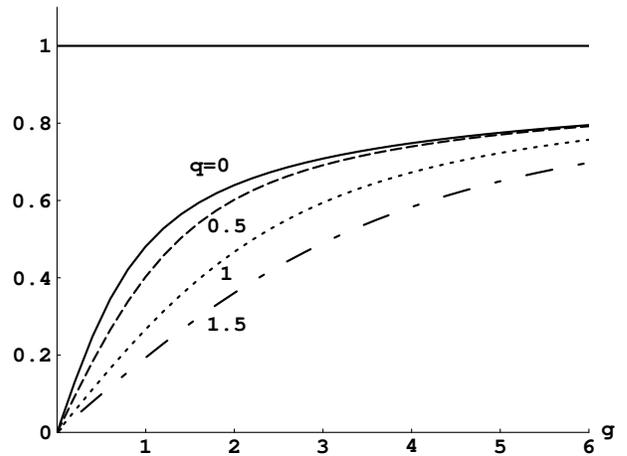}
\caption{Scaling function $\phi_M(g,q)$ for various values of $q$.
Solid line is for $q=0$, dashed line is for $q=0.5$, dotted line is
for $q=1.0$ and dashed chain line is for $q=1.5$.
}
\end{figure}
\end{document}